\documentclass{optica-article}

\journal{opticajournal} 

\articletype{Research Article}

\usepackage{geometry}

\usepackage{graphicx}

\usepackage{lineno}

\begin{document}

\title{Anti-symmetric Multimode Waveguide Grating-Assisted Narrowband MZI for Programmable Spectral Shaping Units}

\author{Qi Wang,\authormark{1} Pin Yu,\authormark{2} Jia Meng,\authormark{1} Jihao Wang,\authormark{1} Zikun Xie,\authormark{1} and Rui Cheng,\authormark{1,*}}

\address{\authormark{1}School of Instrument Science and Opto-Electronics Engineering, Hefei University of Technology, Hefei, Anhui 230009, China\\ \authormark{2}Ningbo University of Technology, Ningbo, 315211, China\\}

\email{\authormark{*}rcheng@hfut.edu.ca} 


\begin{abstract*} 
We present a narrowband integrated Mach-Zehnder interferometer (MZI) capable of precise transmission control within a targeted  wavelength band while maintaining out-of-band transparency. This functionality enables its  use as a fundamental building block for fully programmable on-chip spectral shaping. The device is implemented on a novel dual-mode (TE$_0$/ TE$_1$) transmission platform, where anti-symmetric multimode waveguide Bragg gratings (AM-WBGs) and asymmetric Y-branches are  combined to function as an equivalent narrowband $1\times 2$ or $2\times 2$ coupler. Experimentally, the MZI achieves wide extinction ratio tuning 0 dB to $\sim$30 dB across a $\sim$2.5 nm bandwidth, with independent and simultaneous control of both wavelength and extinction ratio. Cascaded multiple narrowband MZIs are experimentally characterized, demonstrating independent intensity control at individual wavelengths without cross-interference. Furthermore, the device’s application as a tunable, channel-selective optical blocker/passer in high-speed communication systems is experimentally validated. Compared to prior approaches relying on dual-grating-assisted contra-directional couplers, the AM-WBG-based design overcomes fundamental bandwidth limitations caused by unintended intra-waveguide coupling bands. In addition, their single-waveguide-grating structures enhance the reliability of both fabrication and spectral control, while enabling compact spiral configurations for significant miniaturization. These advantages position the proposed MZI a  promising, scalable  candidate for advanced spectral shaping applications.
\end{abstract*}

\section{Introduction}
Over the past decade, silicon-on-insulator (SOI)-based photonic integrated circuits (PICs) have attracted significant research interest due to their high integration density, large bandwidth, and compatibility with mature CMOS fabrication technology \cite{Jalali06}. Among these, programmable PICs—capable of dynamic reconfiguration to enable diverse signal-processing functionalities—have emerged as a rapidly advancing frontier in silicon photonics \cite{bogaerts2020programmable}.
These chips  are now essential  for cutting-edge applications, including quantum photonics \cite{Carolan15, Yu24}, optical computing \cite{Giamougiannis23, Jiang23}, and photonic signal processing \cite{perez2017multipurpose, Hong25}.

A widely adopted approach for programmable PICs  employs  two-dimensional (2D) waveguide meshes or lattices  constructed from Mach-Zehnder interferometer (MZI) unit cells \cite{ capmany2016programmable,  perez2017multipurpose, Perez19, PerezLopez24}. By reconfiguring the circuit topology to control signal interference along different paths, a single chip can realize a wide variety of linear transformations.
However, mesh-based architectures suffer from a non-intuitive relationship between the desired spectral response and the tuning parameters of individual unit cells. As a result, practical deployment often requires extensive Monte Carlo simulations and experimental calibration to construct a functional library of 2D mesh configurations \cite{Perez19}, making the process both time-consuming and costly.
In addition, these architectures typically exhibit periodic spectral responses—e.g., limited by the free spectral range—which inherently constrain their operational bandwidth \cite{Perez16, perez2017multipurpose, PerezLopez24}. Their spectral shaping capabilities are also restricted: while 2D mesh networks can produce standard responses (such as notch, Gaussian, or flat-top filters) and approximate certain target spectra, they struggle to generate arbitrary asymmetric spectral profiles \cite{Zhang20, Zhou20, PerezLopez24}.

To overcome these limitations, Davis et al. proposed a promising programmable PIC building block based on a narrowband MZI functioning as a tunable notch filter \cite{davis2020novel}. This component enables dynamic control of the transmission coefficient within a specific wavelength band while leaving out-of-band wavelengths un-affected. By cascading multiple such filters with different center wavelengths,  arbitrary spectral responses can, in principle, be  programmably achieved, with a direct and intuitive mapping between each unit's tuning parameters and the target spectral response. This straightforward tuning mechanism largely simplifies control, improves reliability, and shows strong potential for advanced spectral manipulation applications.

However, this narrowband MZI implementation relies on 2$\times$2-port grating-assisted contra-directional couplers (CDCs) incorporating closely spaced dual waveguide Bragg gratings. Unlike conventional single-waveguide-based gratings, CDCs suffer from inherent intra-waveguide coupling bands, manifesting as unwanted spectral peaks or notches on either side of the center wavelength \cite{Ikeda08, Tan11, Shi13, Nie19, Xu20, Tunesi25}. These spurious features fundamentally limit  the achievable spectral shaping bandwidth.
Furthermore, the dual-waveguide grating structure introduces fabrication and spectral control complexities. This increased complexity may compromise the practical reliability of spectral performance relative to simpler single-waveguide Bragg gratings.
These drawbacks could become especially problematic when cascading multiple spectral units to construct large-scale programmable photonic circuits. The accumulated limitations may	 ultimately restrict the MZI’s suitability for advanced spectral manipulation applications.

Here, we propose a  silicon photonics narrowband MZI implemented on a novel dual-mode (fundamental $\mathrm{TE_{0}}$ and first-order $\mathrm{TE_{1}}$ transverse electric modes) transmission system to serve as a fundamental building block for  programmable  spectral shaping. The design employs anti-symmetric multimode waveguide Bragg gratings (AM-WBGs) in conjunction with asymmetric Y-junctions to create equivalent narrowband 1×2 or 2×2 couplers. Compared with previous method using grating-assisted CDCs, replacing them with AM-WBGs  overcomes the fundamental bandwidth limitations caused by unwanted intra-waveguide coupling bands.  Furthermore, the single-waveguide-grating structure can enhance the reliability of both fabrication   and spectral control, while enabling compact spiral configurations for significant miniaturization, as extensively validated in previous studies  \cite{Simard13, chen2015spiral, Zou16, Ma18, Cheng18}. 
These combined advantages make the proposed narrowband MZI a promising and scalable solution for arbitrary spectral shaping applications.

\section{PRINCIPLE AND SIMULATIONS}
\subsection{Basic principle}
\begin{figure}[htbp!]
\centering\includegraphics[width=1\linewidth]{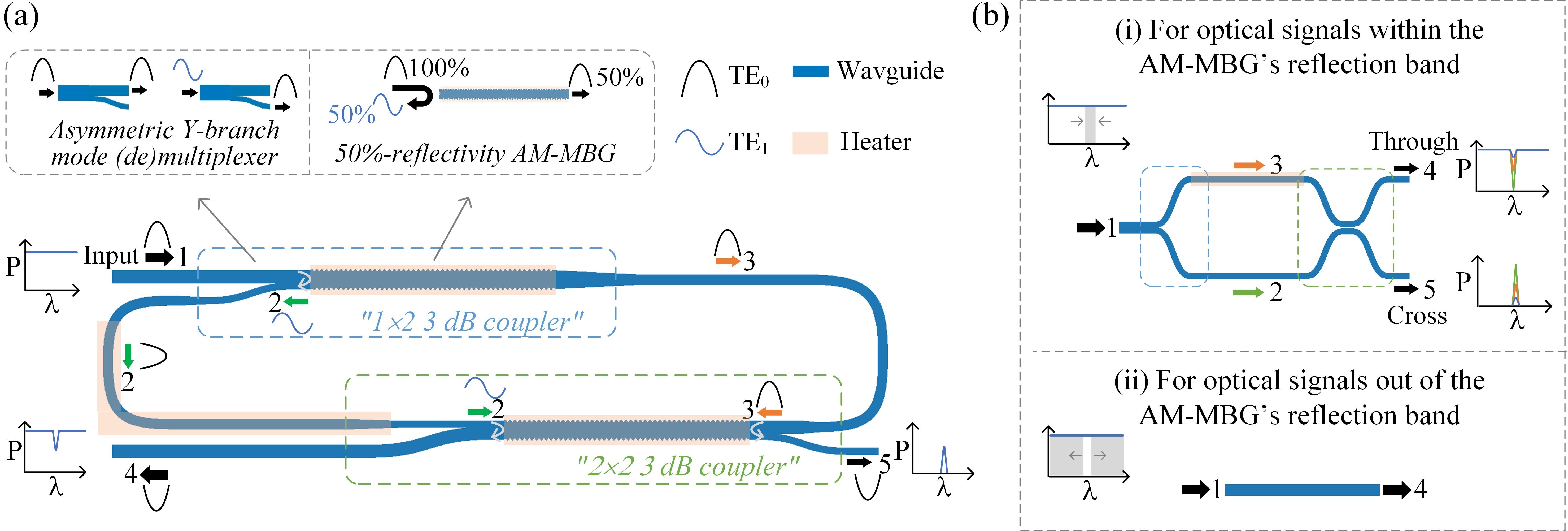}
\caption{ (a) Schematic diagram of the narrowband MZI. (b) Equivalent structure for (i) light transmission within the grating's reflective band and (ii) light transmission outside the grating's reflective band.
}
\label{fig1}
\end{figure}

The narrowband MZI, schematically illustrated in Fig. \ref{fig1}, is implemented on a dual-mode waveguide system supporting both $\mathrm{TE_{0}}$ and $\mathrm{TE_{1}}$ modes. This configuration functions as an interferometer only within a specific narrow spectral range, corresponding to the operating bandwidth of the waveguide grating [Fig. \ref{fig1}(b)-(i)]. For wavelengths outside this band, light bypasses the interferometer entirely and exits directly through the straight-through port, effectively behaving as a simple waveguide [Fig. \ref{fig1}(b)-(ii)].
The relative phase difference between the interferometer arms can be tuned via thermo-optic or electro-optic effects. Consequently, this system allows for dynamically controlling transmission within the target wavelength band while maintaining transparency at out-of-band wavelengths.
The output port through which out-of-band signals exit is termed the "through" port (Port 4 in Fig. \ref{fig1}), while the complementary output is termed the "cross" port [Port 5 in Fig. \ref{fig1}]. As a result, the through port exhibits a tunable narrowband notch response, whereas the cross port provides a controllable bandpass characteristic.

A direct application of this system is as a tunable notch or bandpass filter. More importantly, this architecture can serve as a fundamental spectral shaping unit. 
By cascading multiple such spectral control units—each with a linearly varying center wavelength—through their through ports and dynamically controlling the transmission distribution across different discrete wavelengths, the spectral response of the cascaded system in principle can be reconfigured into arbitrary shapes. This capability facilitates the implementation of arbitrarily programmable filters.

Below, we will elaborate in more details of  the operation principle of the system. The structure employs two identical AM-WBGs to replace traditional broadband directional couplers, and combines them with three asymmetric Y-branches to form a compact ring structure. The AM-WBGs function as narrowband mode-converter-reflector with a 3 dB (50\%) reflectivity. Specifically, they are designed to have the following features.
	\begin{itemize}
		\item When the incident light wavelength is within the grating's operating wavelength range, if the incident light mode is TE$_0$ or TE$_1$, the grating will reflect 50\% of the incident energy and convert it to TE$_1$ or TE$_0$ mode, respectively.  The remaining 50\% of the energy is transmitted through the grating.
		\item When the incident light wavelength is outside the grating's operating range, the light signal passes through the grating without loss.
	\end{itemize}

The asymmetric Y-branches are designed as broadband TE\textsubscript{0} and TE\textsubscript{1} mode multiplexers (demultiplexers). More specifically:
\begin{itemize}
	\item When TE\textsubscript{0} and TE\textsubscript{1} modes are incident from the common waveguide of the Y-branch, they are output as TE\textsubscript{0} modes from the wider and narrower arms, respectively.
	
	\item Conversely, when TE\textsubscript{0} modes are input from the wider and narrower arms, they are converted to TE\textsubscript{0} and TE\textsubscript{1} modes, respectively, and both are output from the common waveguide.
\end{itemize}

Based on the  characteristics of the AM-WBGs and asymmetric Y-branches, they can be combined to form an equivalent narrowband "1$\times$2" or "2$\times$2" 3 dB coupler, as enclosed by the blue and green dashed boxes in Fig. \ref{fig1}(a), respectively. By connecting the two output ports of the 1$\times$2 coupler to the input ports of the 2$\times$2 coupler through waveguides, a narrowband 1$\times$2 MZI is realized. In this configuration, the interconnected waveguides function as the MZI’s interference arms, with Beams 2 and 3 [labeled in Fig. \ref{fig1}(a)] acting as the interfering beams. By incorporating a thermal phase shifter into one of the arms, the phase difference between the two paths can be dynamically adjusted, enabling precise tuning of the narrowband MZI response.

\subsection{AM-WBG}
\begin{figure}[ht!]
\centering\includegraphics[width=7.6cm]{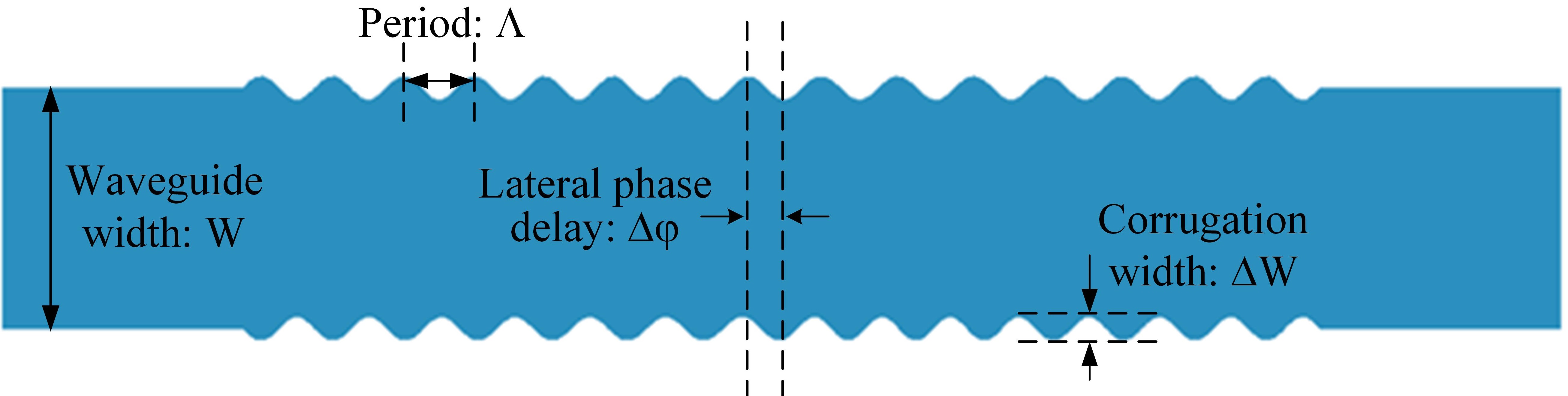}
\caption{Schematic structure of an AM-WBG.
}
\label{fig2}
\end{figure}

The AM-WBG designed in this work, as schematically illustrated in Fig. 2, is a sidewall-modulated multimode waveguide Bragg gratings where the two sides of the corrugations are laterally misaligned. Such a waveguide Bragg grating, due to its antisymmetric structural perturbation, can allows contra-directional mode coupling between an even-order and an odd-order mode \cite{Castro05}, which here are $\mathrm{TE_{1}}$ and $\mathrm{TE_{0}}$ modes. The AM-WBGs required by the narrowband MZI should satisfy that 1) the center wavelength matches the wavelength of the MZI, 2) the spectral sidelobes are small, 3) the reflection bandwidth meets the designed value, and 4) the maximum reflectivity is equal $50\%$. Here, wavelength is set to $\sim $1550 nm, and the bandwidths are chosen to $\sim $2 nm.

To design the AM-WBG meeting  the requirements above, first the waveguide's width should be chosen to ensure it  supports both $\mathrm{TE_{0}}$ and $\mathrm{TE_{1}}$ modes. Then, the grating period, $\Lambda$,  should be selected such that the following phase-matching condition can be satisfied at the designed wavelength:
\begin{equation}
n_{0} +  n_{1} = \frac{\lambda }{\Lambda }    
\end{equation}
where $n_{0}$ and  $n_{1}$ are the effective indices of the $\mathrm{TE_{0}}$ and $\mathrm{TE_{1}}$ modes, respectively.

\begin{figure}[htbp]
	\centering\includegraphics[width=5.6cm]{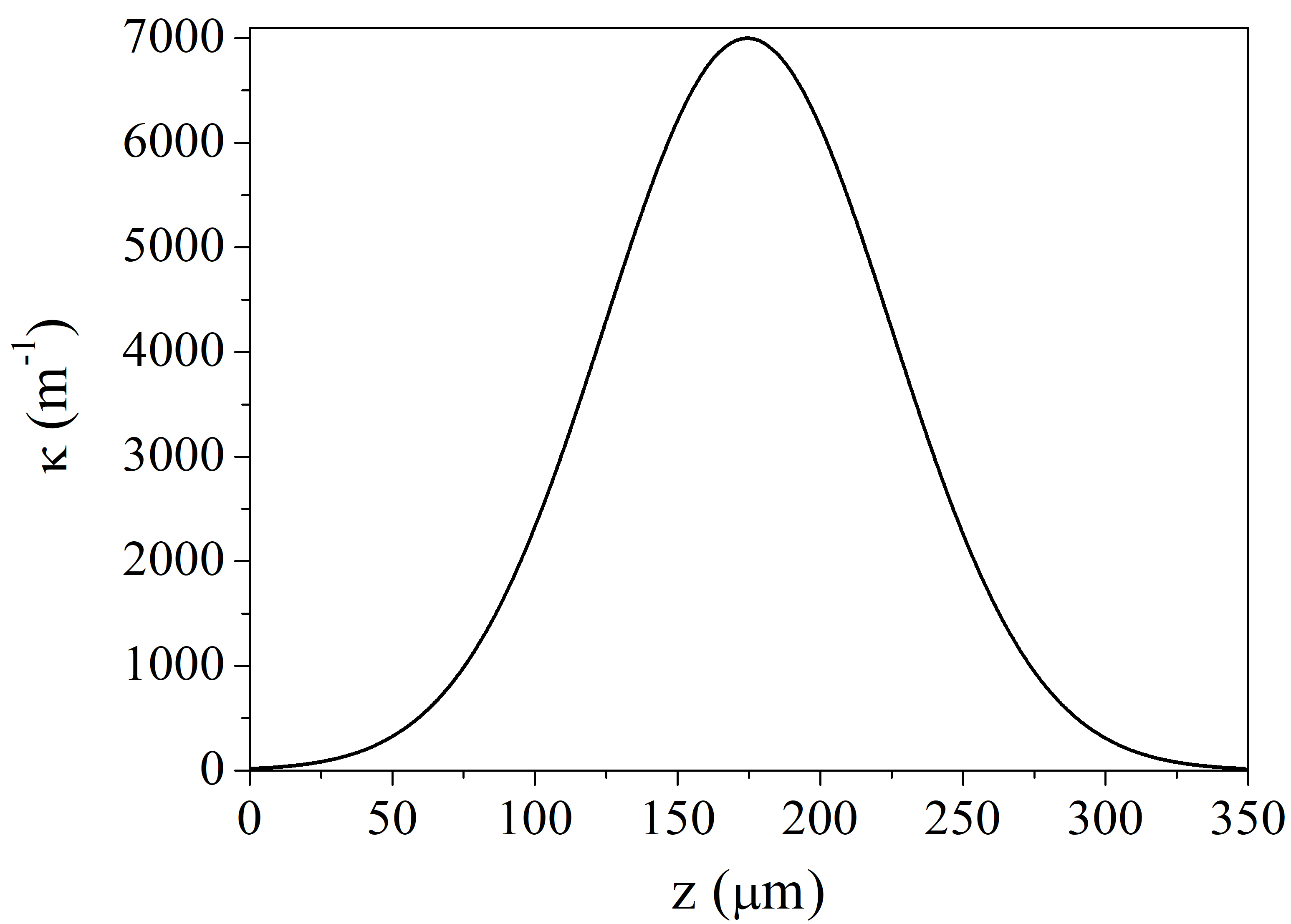}
	\caption{Gaussian apodization profile of the AM-WBG.
	}
	\label{fig3}
\end{figure}

A Gaussian apodization is applied to the AM-WBG to significantly suppress the side-lobes. The coupling strength 
$\kappa$ as a function of the grating position 
$z$ follows the profile:
\begin{equation}
\kappa (z) =  
\kappa _{\text{max}} \exp\left(-4 \ln(2) \left( \frac{z - L/2}{L/2.95} \right)^2 \right)
\end{equation} \label{kappaZ}
where  $\kappa _{\text{max}}$  is the maximum coupling strength  along the grating, which can be controlled by  the corrugation width, $\Delta W$,  and $L$ is the total length of the grating.
The apodization is performed by modulating the lateral phase delay [denoted as $\Delta \varphi $] between the two sides of the corrugation \cite{Cheng21}. 
For the AM-WBG, $\kappa $ between $\mathrm{TE_{0}}$ and $\mathrm{TE_{1}}$ will be related to $\Delta \varphi $ via the following equation:
\begin{equation}
\kappa \propto \sin \left ( \frac{\Delta \varphi }{2}  \right )    
\end{equation} \label{kappaPhi}
The lateral phase delay $\Delta \varphi$ along the grating, denoted as $\Delta \varphi(z)$, can be calculated from the coupling strength profile $\kappa(z)$. The resulting $\Delta \varphi(z)$  then can be used to construct the full grating structure \cite{Cheng21}.

\begin{figure}[htbp!]
\centering\includegraphics[width=11cm]{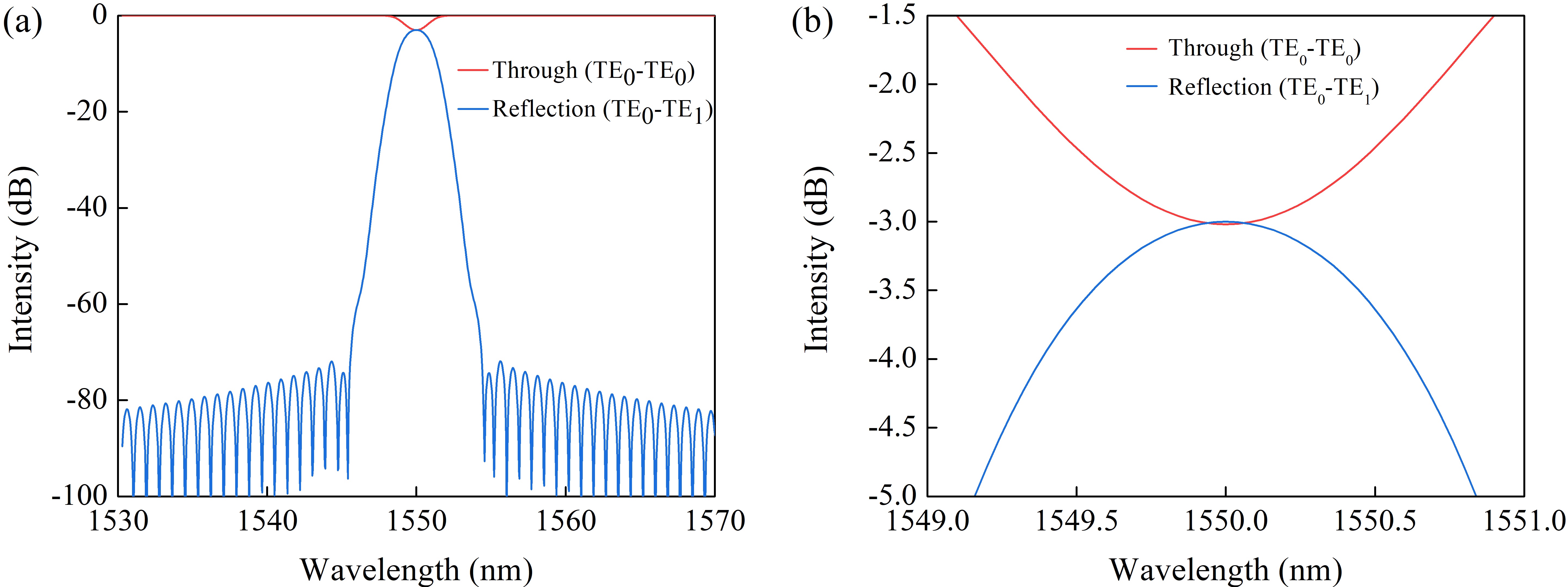}
\caption{ Simulated spectral response of the designed AM-WBG. (a) Full spectral range. (b) Magnified view showing the peak reflection region.
}
\label{fig4}
\end{figure}

For the finally designed AM-WBG, the waveguide width, $W$, is 1 $\upmu$m, $L$ is 352.5 $\upmu$m, $\Lambda$ is 300 nm, and $\Delta W$ is 16 nm, leading to $\kappa_{\text{max}}$ of about 7000 m$^{-1}$ . 
The Gaussian apodization profile of the designed AM-WBG is plotted in Fig. \ref{fig3}. The reflection and transmission  responses of the AM-WBG, calculated using the coupled-mode theory (CMT) based transfer matrix method (TMM), are presented in Fig. \ref{fig4}. It can be seen that the maximum reflectivity is -3 dB, the 3 dB bandwidth is around 2 nm, and the side-lobe suppression ratio owning to the Gaussian apodization is as high as 69 dB.

\subsection{Asymmetric Y-branch}
\begin{figure}[htbp]
\centering\includegraphics[width=9cm]{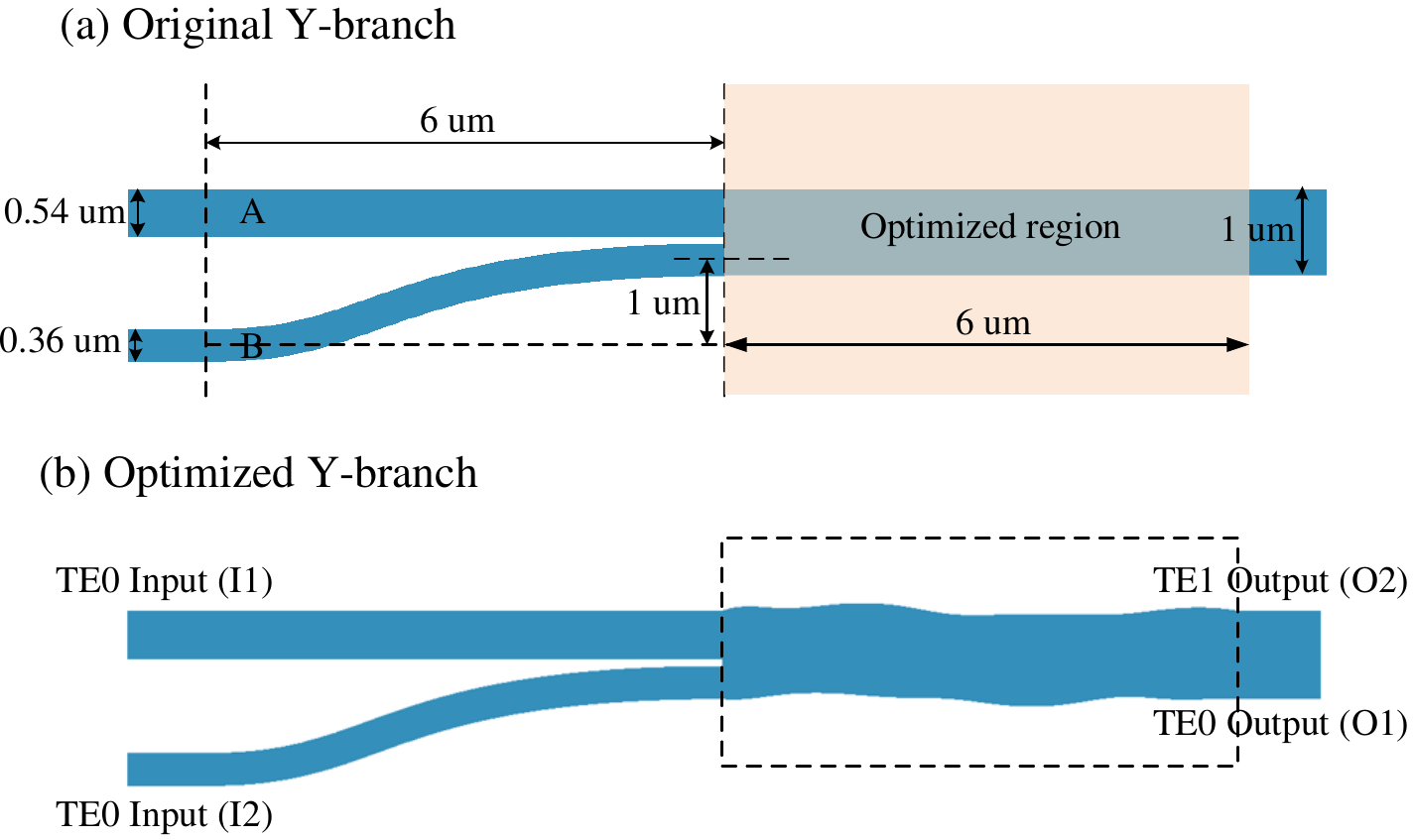}
\caption{ Schematic illustrations of (a) the original and (b) optimized asymmetric Y-branches.
}
\label{fig5}
\end{figure}

In the proposed narrowband MZI structure, asymmetric Y-branches are utilized as a mode-dependent router. 
Specifically, when proper waveguide widths are employed, the input $\mathrm{TE}_{0}$ and $\mathrm{TE}_{1}$ modes from the common waveguide can be adiabatically routed to the $\mathrm{TE}_{0}$ modes of the wider and narrower output arms, respectively \cite{Love12}.
For optimal performance, the Y-branches should ideally exhibit low crosstalk, small footprint, and wide operational bandwidth. However, conventional asymmetric Y-branches often fall short in achieving these desired characteristics simultaneously. To address this challenge, we employ the inverse design  to optimize the asymmetric Y-branches \cite{wang2023inverse}.

The design process begins by defining the basic Y-branch structure, where one arm is configured as an S-shaped curve while the other remains straight [Fig.~\ref{fig5}(a)]. This asymmetric configuration facilitates the integration of the MZI loop. 
The Y-branch arms are designed with distinct widths: 0.54~$\upmu$m for  Arm A and 0.36~$\upmu$m for Arm B. 
The S-shaped Arm B has horizontal and vertical lengths of 6 $\upmu$m and 1 $\upmu$m, respectively. 
The multimode common waveguide features a 1~$\upmu$m width to ensure  support for the $\mathrm{TE_{0}}$ and $\mathrm{TE_{1}}$ modes.  
Through iterative optimization of the  6 $\upmu$m-long common waveguide’s shape, we minimize mode crosstalk and insertion loss while maintaining an ultra-compact footprint.

The figure of merit (FOM) for the optimization process is defined as the combined sum of (i) the $\mathrm{TE}_0$ mode power at the wider arm  under  $\mathrm{TE}_0$  excitation  at the common waveguide and (ii) the $\mathrm{TE}_0$ mode power at the narrower arm under  $\mathrm{TE}_1$  excitation  at the common waveguide.
The adjoint method is used to efficiently compute the gradient information necessary for optimization  \cite{LalauKeraly13}. The optimization is carried out using a three-dimensional finite-difference time-domain (3D-FDTD) method, achieving convergence after 40 iterations. The optimized Y-junction waveguide structure is depicted in Fig. \ref{fig5}(b).

\begin{figure}[ht!]
\centering\includegraphics[width=9cm]{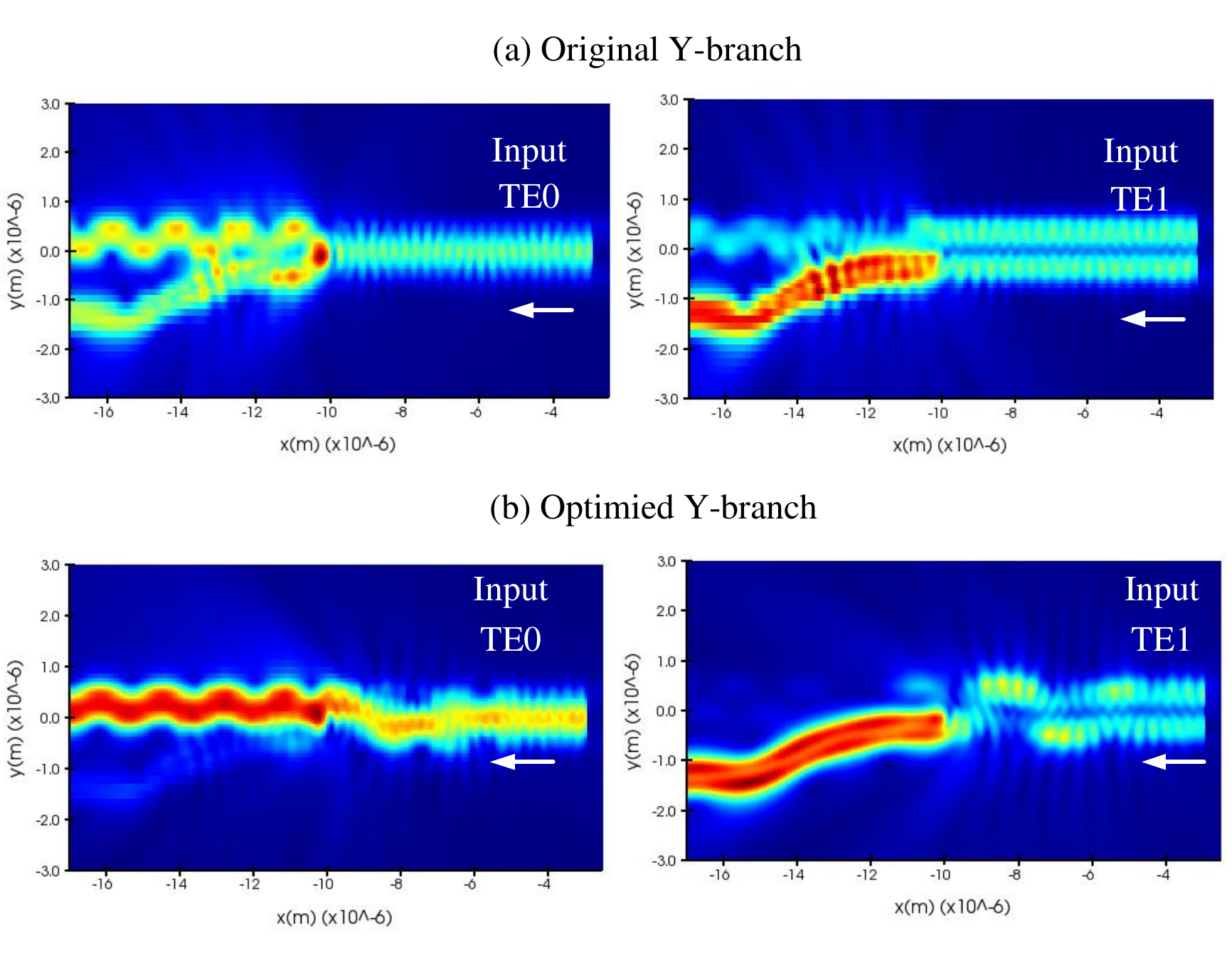}
\caption{ Comparison of electrical field distributions of (a) the original and (b) optimized asymmetric Y-branches under  TE$_0$/TE$_1$ mode excitation from the common waveguide.
}
\label{fig6}
\end{figure}

\begin{figure}[htbp]
\centering\includegraphics[width=10.6cm]{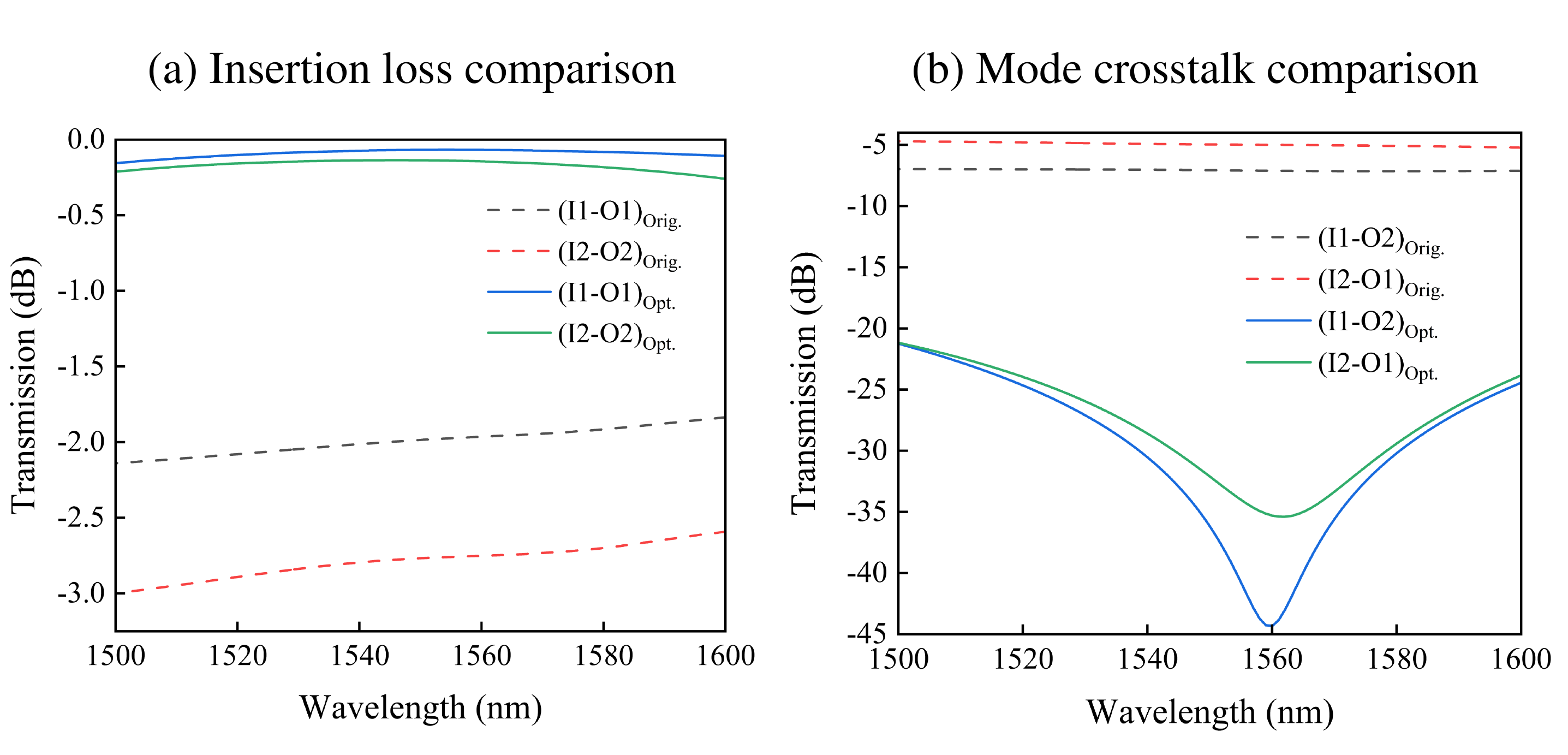}
\caption{ Performance comparison between the original (dashed) and optimized (solid) asymmetric Y-branches.
}
\label{fig7}
\end{figure}

The optimized Y-branch is characterized using the 3D-FDTD method. Figure \ref{fig6} compares the electric field intensity distributions of the initial and optimized Y-branches when inputting the different modes from the common waveguide, at the wavelength 1550 nm. Significant crosstalk can be seen in the original Y-branch for each input mode, while for the optimized one, no appreciable crosstalk is observed, demonstrating the large improvement in mode (de)multiplexing performance of the asymmetric Y-branch brought by the optimization. More specifically, the transmission efficiency of the $\mathrm{TE_{0}}$ mode into the wider arm is increased from $63\%$ to $97.8\%$, while that of the $\mathrm{TE_{1}}$ mode into the narrower arm improved from $53\%$ to $96.9\%$ due to the optimization. 
Figure \ref{fig7}  compare the calculated insertion loss and mode crosstalk, respectively, of the original and optimized Y-branches over a wavelength range of 1500-1600 nm. A significant decrease in both insertion loss and mode crosstalk over the characterized wavelength range owing to the optimization is observed. Specifically, the insertion losses for the $\mathrm{TE_{0}}$ and $\mathrm{TE_{1}}$ modes of the Y-branch within the simulated spectral range are reduced from $<$2.13 dB to $<$0.15 dB and from $<$3 dB to $<$0.26 dB, respectively, while the mode crosstalks for the $\mathrm{TE_{0}}$ and $\mathrm{TE_{1}}$ modes are decreased from $<$-7 dB to $<$-21.2 dB and from $<$-4.7 dB to $<$-21.2 dB, respectively.

\subsection{Narrowband MZIs}
\subsubsection{Single MZI}
\begin{figure}[ht!]
\centering\includegraphics[width=10cm]{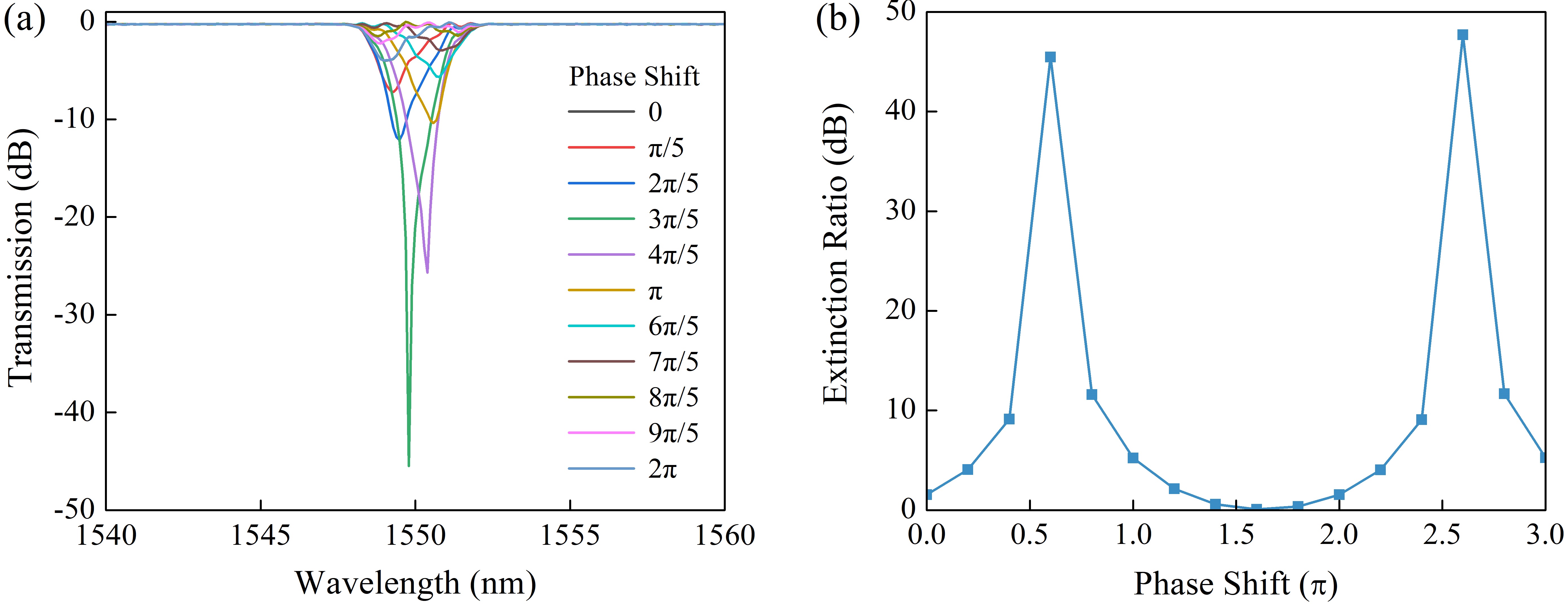}
\caption{Simulated performance of the narrowband MZI. (a) Spectral response evolution under  phase tuning. (b) Extinction ratio as a function of phase shift.
}
\label{fig8}
\end{figure}

The narrowband MZI is constructed using the above-designed  AM-WBG and asymmetric Y-branch. The MZI is  simulated using Lumerical Interconnect,  a system-level photonic integrated circuit simulator. In the system level, it is found that the waveguide length should be carefully selected to balance the optical path difference between the two arms of MZI. Otherwise, in addition to the extinction ratio, unintended wavelength shift of the narrowband MZI could also happen when tuning the phase shifter due to the interference pattern drift. Figure \ref{fig8}(a) presents the simulated transmission response as a function of the applied phase shift.
The notch depth varies with the phase shift, while the out-of-band response remains unaffected, demonstrating the MZI’s capability as a spectral shaping unit for selective wavelength intensity control.
Additionally, Fig. \ref{fig8}(b) plots the extinction ratio versus the phase shift. The extinction ratio follows a sinusoidal trend, tunable from 0 dB to 45.5 dB. 

\subsubsection{Cascaded MZIs}
\begin{figure}[ht!]
\centering\includegraphics[width=13.2cm]{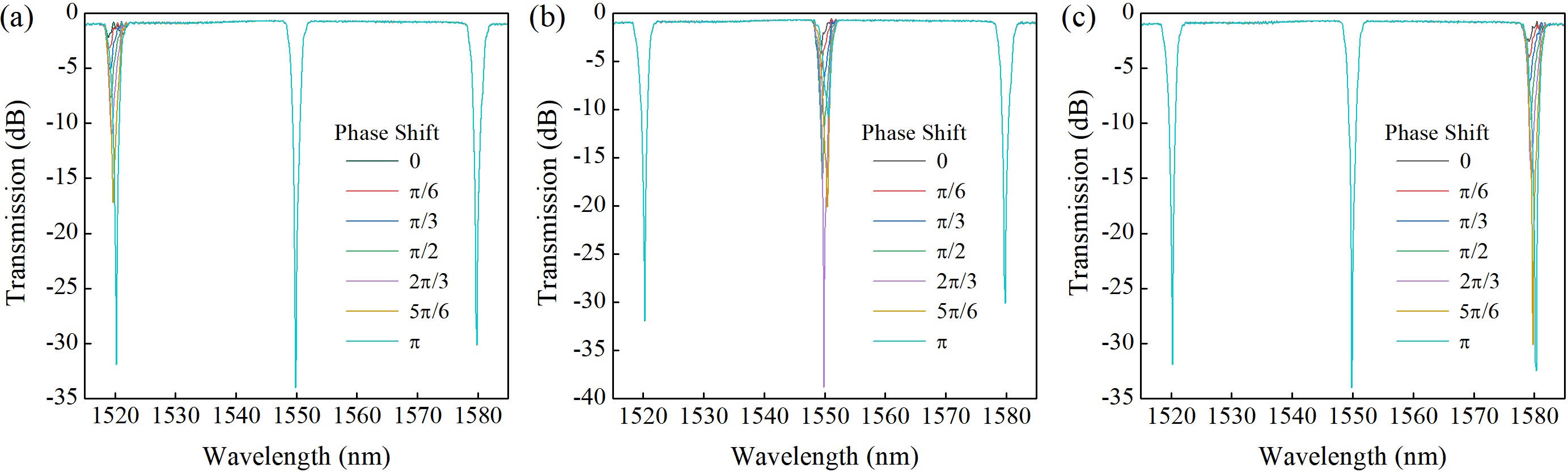}
\caption{Simulated spectral response evolution of the cascaded narrowband MZI system while independently tuning the (a) 1520 nm, (b) 1550 nm, and (c) 1580 nm MZI stages.
}
\label{fig9}
\end{figure}

Cascading multiple narrowband MZIs with different center wavelengths can allow  transmissions at multiple wavelengths to be individually controlled, which can potentially achieve arbitrary spectral shaping. To demonstrate this capability, three MZIs with center wavelengths of 1520 nm, 1550 nm, and 1580 nm are cascaded and simulated, using Interconnect.
Each MZI is individually controlled, and the combined system response is calculated.
Figures \ref{fig9}(a)–\ref{fig9}(c) show the spectral evolution when tuning the 1520 nm, 1550 nm, and 1580 nm MZIs, respectively. Notably, only the transmission at the wavelength corresponding to the tuned MZI is modified, while the others remain unaffected. This confirms the feasibility of achieving arbitrary spectral shaping using cascaded MZI units.

\subsubsection{Impact of the AM-WBG wavelength misalignment on the system performance}
\begin{figure}[ht!]
\centering\includegraphics[width=10.6cm]{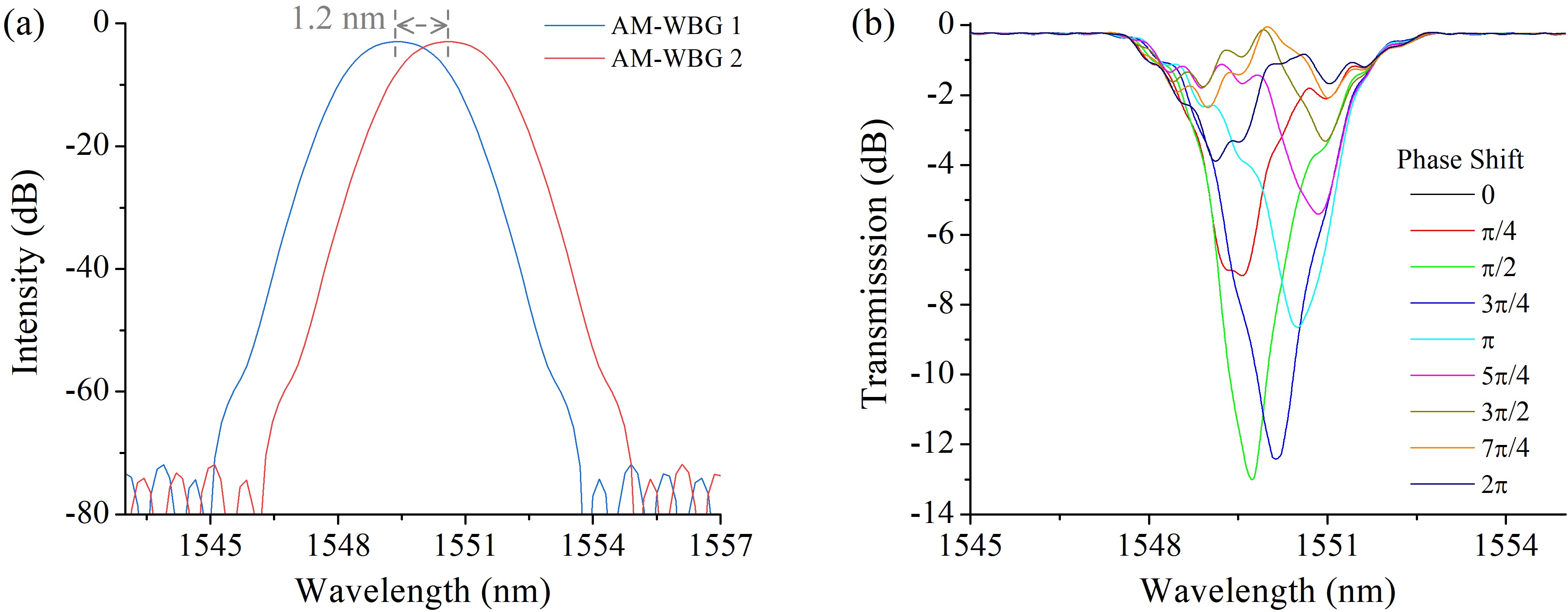}
\caption{
Investigation of wavelength misalignment  in AM-WBGs on narrowband MZI performance.
(a) Reflection spectra of the AM-WBGs.
(b) Evolution of the MZI transmission response under the phase tuning.
}
\label{fig10}
\end{figure}

Due to potential manufacturing errors, the central wavelengths of the two AM-WBGs may become misaligned. To assess the impact of this misalignment on system performance, we simulate the spectral response of a narrowband  MZI  while intentionally introducing a wavelength offset of 1.2 nm between the two AM-WBGs [Fig. \ref{fig10}(a)].
The simulation results [Fig. \ref{fig10}(b)] show that the response exhibits two small notches even when the MZI is tuned to its flattest state. Additionally, the maximum achievable extinction ratio is reduced to $\sim$13 dB, significantly lower than that of the aligned case [45.5 dB, as shown in Fig. \ref{fig8}(b)]. These results indicate that spectral misalignment between the AM-WBGs can degrade both spectral shaping performance and the extinction ratio tuning range.

\section{Experimental results}
\subsection{Fabrication and testing setup}
\begin{figure}[ht!]
\centering\includegraphics[width=13cm]{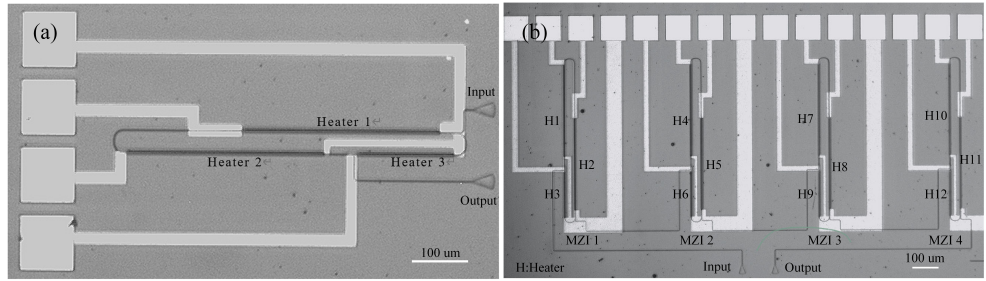}
\caption{Optical microscope images of  fabricated  (a) single and (b) cascaded narrowband MZIs. 
}
\label{fig11}
\end{figure}

The devices were fabricated using electron-beam lithography on a standard commercial SOI wafer. The wafer consists of a 3 um thick buried silicon dioxide layer and a 220 nm thick silicon layer. Ti/W alloy heaters were deposited above the MZI arm and AM-WBGs for thermal tuning. Figures \ref{fig11} (a) and \ref{fig11}(b) show the optical microscope images of the fabricated single and cascaded MZIs, respectively. 
A tunable laser source (Keysight 81960A) and a multi-channel  optical power meter (Agilent N7745A) were used to characterize the spectra of the devices. A high-precision programmable voltage source (Keysight E36312A) was employed for the electrical tuning of the micro-heaters.

\subsection{AM-WBG}
\begin{figure}[ht!]
\centering\includegraphics[width=5.5cm]{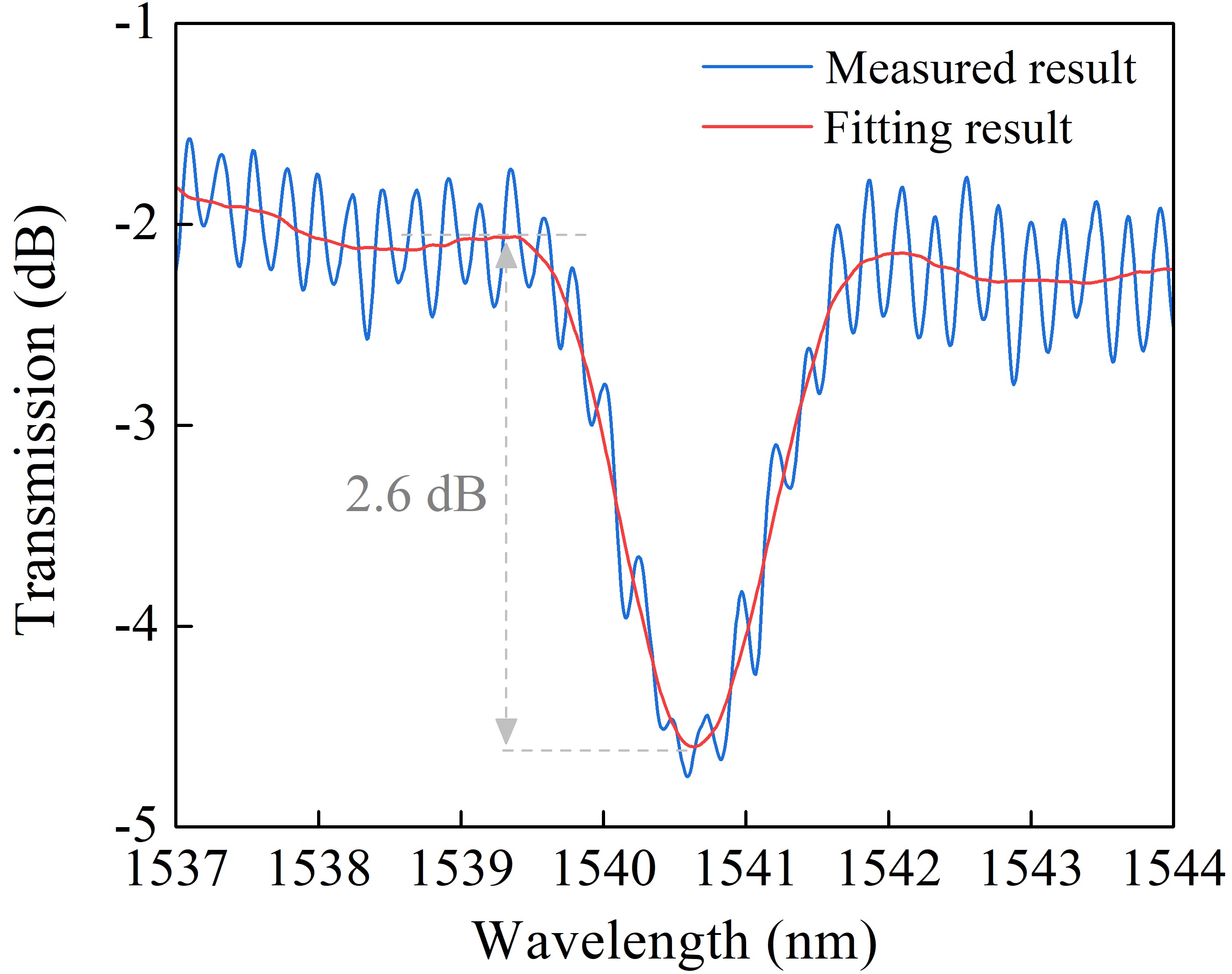}
\caption{Measured through-port response of the AM-WBG.
}
\label{fig12}
\end{figure}

The individual AM-WBG was fabricated and tested. 
The  AM-WBG  maintains the design parameters outlined in Section 2.2. 
Figure \ref{fig12} exhibits the normalized transmission spectra of the AM-WBGs, revealing a distinct transmission notch at a center wavelength of 1540.7 nm. This notch arises from contra-directional coupling between the TE$_0$ and TE$_1$ modes.  The measured notch depth of 2.6 dB suggests that the fabricated device achieves a maximum reflectivity approaching the ideal 50\% target.

\subsection{Asymmetric Y-branch}
\begin{figure}[ht!]
\centering\includegraphics[width=10.3cm]{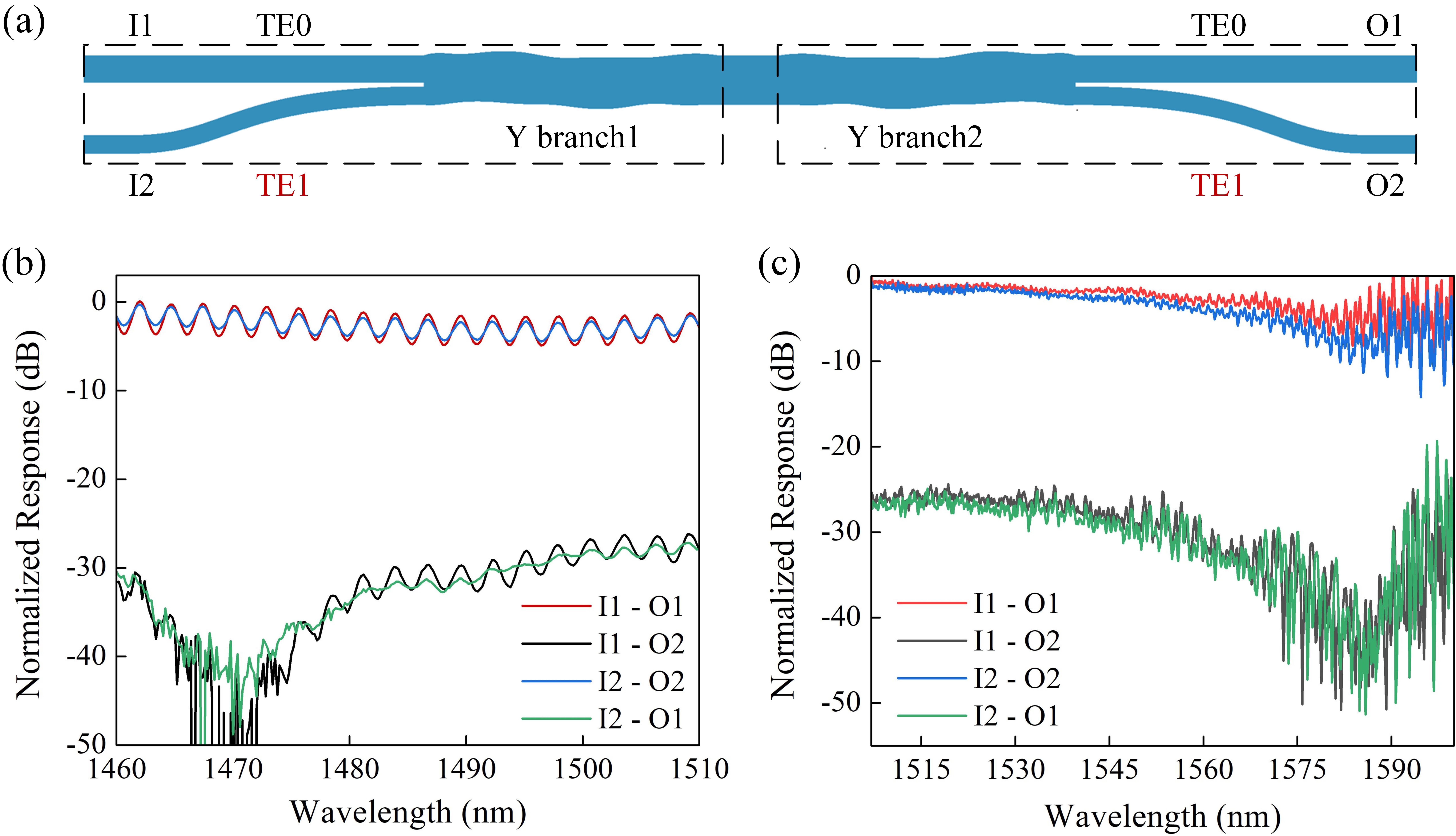}
\caption{(a) Schematic of the asymmetric Y-branch MDM system.
	(b) Measured transmission responses over (b) 1460–1510 nm and (c) 1507–1600 nm wavelength ranges.
}
\label{fig13}
\end{figure}

To experimentally characterize the inverse-designed asymmetric Y-branch, we fabricated and measured mode division multiplexing (MDM) systems composed of back-to-back identical but oppositely oriented Y-branches [Fig. \ref{fig13}(a)].
Due to the limited 1507–1630 nm tuning range of our tunable laser (Keysight 81960A), measurements were performed using two complementary setups. The transmission response across 1507–1600 nm was acquired using the tunable laser, while the 1460–1510 nm spectral range was characterized using a broadband light source (WL-SC400-2-PP) and an optical spectrum analyzer (Yokogawa AQ6370C).
Figures \ref{fig13}(b) and \ref{fig13}(c) present the measured transmission responses of the MDM circuits incorporating the optimized asymmetric Y-branches for these respective wavelength bands. 
The fabricated devices exhibit mode crosstalk below $-23\,\mathrm{dB}$ across a broad wavelength range from $1460\,\mathrm{nm}$ to $1590\,\mathrm{nm}$. The measured insertion losses at the center wavelength of $1525\,\mathrm{nm}$ were approximately $1.5\,\mathrm{dB}$ for both the $\mathrm{TE}_0$ and $\mathrm{TE}_1$ modes. These results demonstrate the high performance of the inverse-designed Y-branches, confirming their suitability for implementation in narrowband MZIs.

\subsection{Single narrowband MZI}
\begin{figure}[ht!]
\centering\includegraphics[width=11cm]{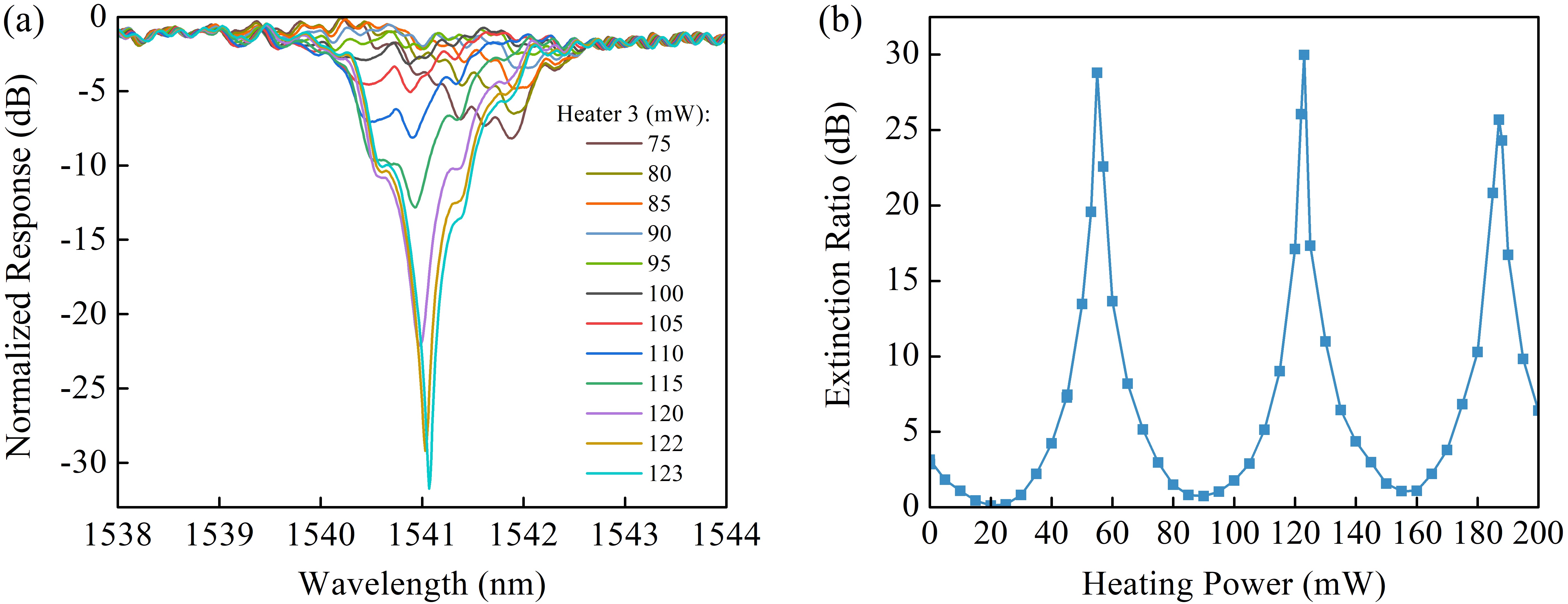}
\caption{(a) Evolution of the transmission response as the thermal phase shifter is electrically tuned. (b) Extinction ratio  as a function of heating power.
}
\label{fig14}
\end{figure}

The narrowband MZI was experimentally characterized. A microheater (Heater 3) was integrated along a 180-$\upmu$m section of one interferometer arm, while two additional microheaters (Heaters 1 and 2) were positioned over both  AM-WBGs.
The spectral response of the MZI to applied heating power was measured, as illustrated in Fig. \ref{fig14}(a). 
The transmission near the MZI's center waveguide ($\sim$1541.1 nm) is significantly tuned with applied heating power over a bandwidth of $\sim$2.5 nm.
Figure \ref{fig14}(b) presents the extinction ratio as a function of electrical power, demonstrating a sinusoidal variation that closely matches the theoretical prediction shown in Fig. \ref{fig8}(b). The extinction ratio is continuously tunable from 0 dB up to about 30 dB, with an electrical power of approximately 35 mW required to transition between the maximum and minimum extinction states.

\begin{figure}[ht!]
\centering\includegraphics[width=5.8cm]{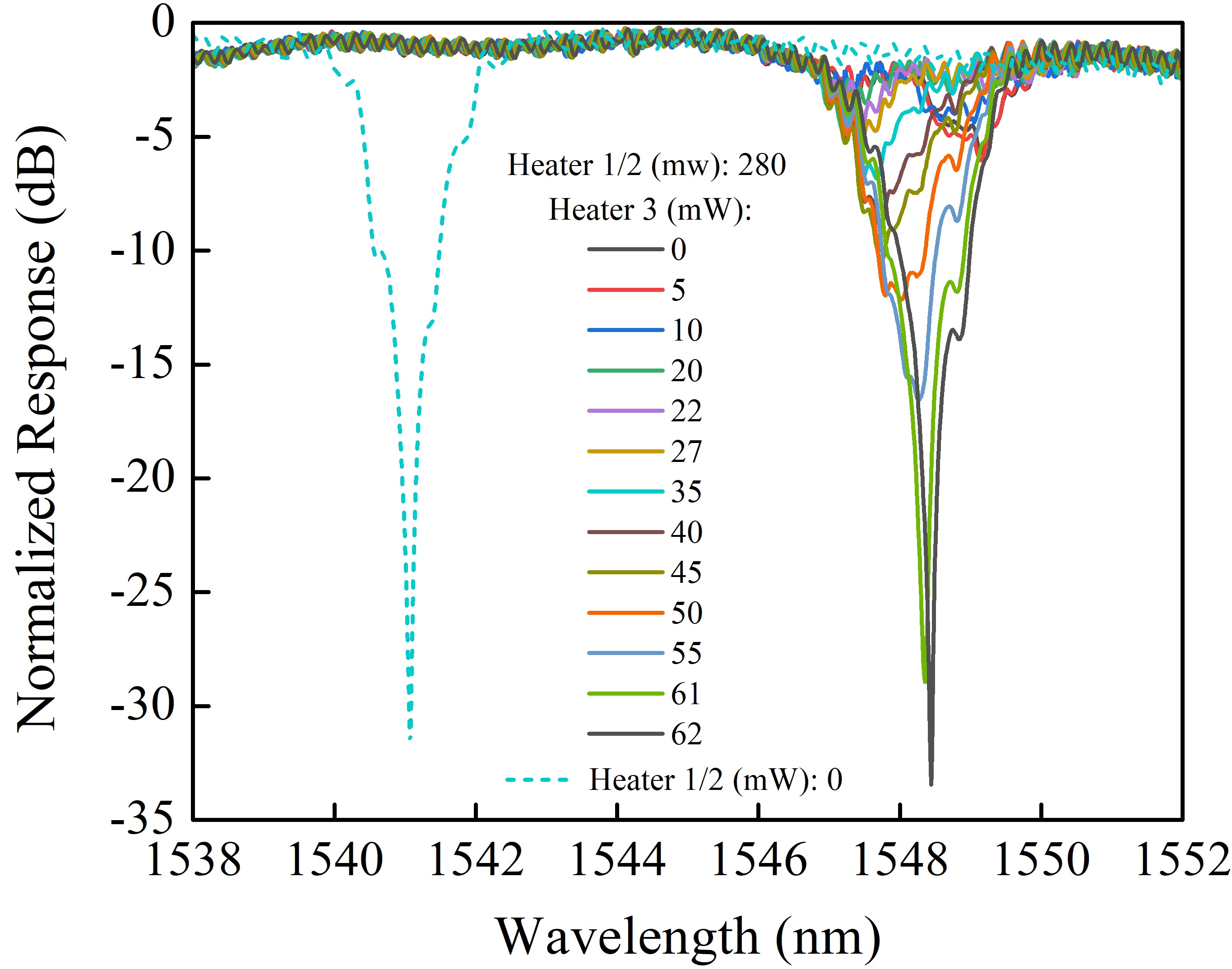}
\caption{
	Extinction ratio tuning  of the narrowband MZI as the wavelength is adjusted from $\sim$1541 nm to $\sim$1548.4 nm.
}
\label{fig15}
\end{figure}

Next, we further investigated the ability to simultaneously tune both the wavelength and extinction ratio of the narrowband MZI. In the experiment, the heaters above both AM-WBGs (Heater 1 and Heater 2) were adjusted under a fixed electrical power of 280 mW to shift the center wavelength of the narrowband MZI. Subsequently, the extinction ratio was fine-tuned by gradually varying the electrical power applied to Heater 3. The spectral responses under different Heater 3 power levels are shown in Fig. \ref{fig15}. The wavelength shifted from $\sim$1541 nm to $\sim$1548.4 nm, while the extinction ratio was adjusted at the new wavelength. These results demonstrate the successful reconfiguration of both  wavelength and extinction ratio in the narrowband MZI.

\subsection{Cascaded  narrowband MZIs}
\begin{figure}[htbp!]
\centering\includegraphics[width=10.6cm]{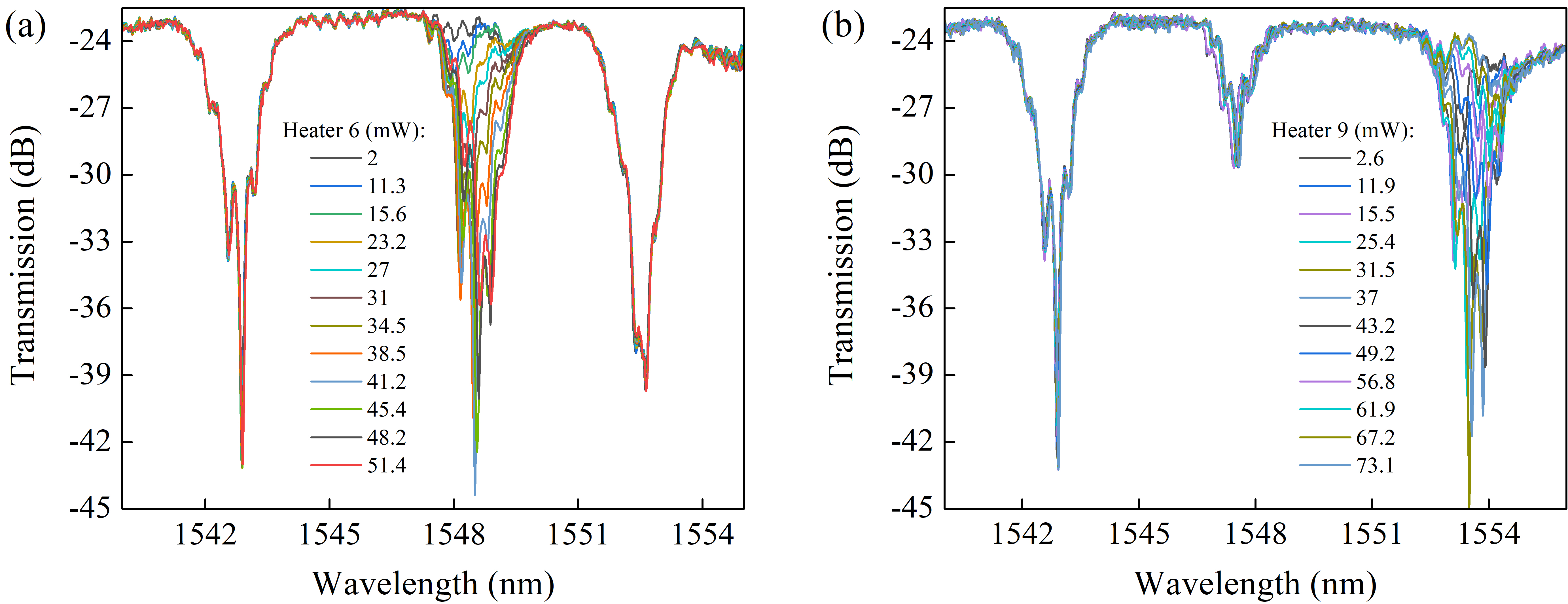}
\caption{Measured spectral evolution in cascaded narrowband MZIs under individual  tuning of (a) the $\sim$1548\,nm MZI and (b) the $\sim$1553\,nm MZI, demonstrating independent wavelength control in the cascaded system.
}
\label{fig16}
\end{figure}

Cascading  narrowband MZIs with different center wavelengths enables to control the transmissions at different wavelengths, which potentially allows arbitrary spectral response shaping. To demonstrate this  potential, we experimentally characterized a fabricated system consisting of four identical cascaded narrowband MZIs.
The center wavelengths of the MZIs were tuned independently to separate their spectral responses. Due to experimental constraints, only two MZIs  were wavelength-tuned. Initially, all MZIs had a center wavelength of $\sim$1543 nm; two were then electrically tuned to $\sim$1548 nm and $\sim$1553 nm, respectively. Following wavelength separation, the extinction ratios of the MZIs were individually adjusted.
Figures \ref{fig16}(a) and \ref{fig16}(b) show the tuning results for the MZIs at $\sim$1548 nm and $\sim$1553 nm, respectively. Only the intensity at the tuned MZI's wavelength was modified, while other wavelengths remained unaffected. 
These results confirm independent wavelength control and validate the potential of arbitrary spectral shaping in the cascaded system.

\subsection{Tunable, channel-selective  blocker/passer in  high-speed optical communications}

One immediate application of the narrowband MZI is its use as an extinction-ratio-tunable notch filter, which can serve as a reconfigurable channel-selective attenuator or blocker/passer in optical communication systems. To validate the latter functionality, an experiment was conducted to measure the eye diagram of a transmission link incorporating the single narrowband MZI (previously characterized in Section 3.4). 

\begin{figure}[htbp!]
	\centering\includegraphics[width=10.6cm]{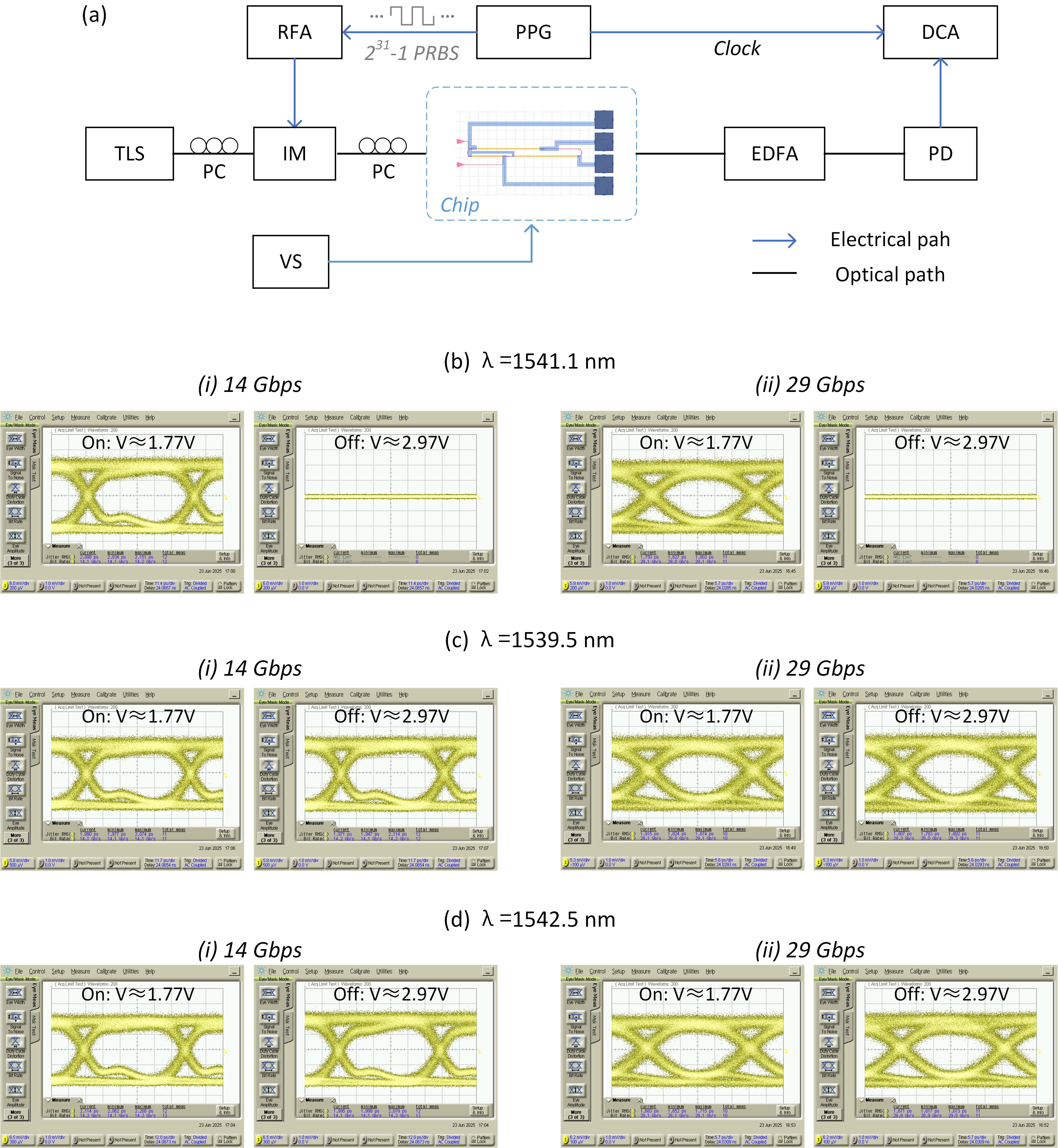}
	\caption{(a) Experimental setup of the  high-speed optical transmission test; 
		(b) Optical eye diagrams at the narrowband MZI center wavelength (1541.1 nm), with the MZI tuned to its "on" and "off" states, for PRBS data rates of (i) 14 Gbps and (ii) 29 Gbps.
		(c) and (d) Corresponding eye diagrams at wavelengths outside the MZI's bandwidth: (c) 1539.5 nm and (d) 1542.5 nm.
	}
	\label{fig.eye}
\end{figure}

The experimental setup is illustrated in Fig. \ref{fig.eye}(a).
The tunable laser (TLS) acted as the system's light source. 
The laser output passed through a polarization controller (PC) before being fed into a commercial LiNbO$_3$ MZI intensity modulator (IM, Fujitsu FTM7938EZ). The IM was driven by a pulse pattern generator (PPG)  producing a non-return-to-zero (NRZ) 
$2^{31}-1$ pseudo-random binary sequence (PRBS). Prior to modulation, the PRBS signal was amplified by a 40 GHz RF amplifier (RFA).
The modulated optical signal from the IM passed through a second PC, was coupled into the integrated narrowband MZI, and then output into an optical fiber via vertical grating couplers.  The light was subsequently amplified by an erbium-doped fiber amplifier (EDFA) and converted to an electrical signal using a 40 GHz high-speed photodetector (PD, Finisar XPDV2120RA). The PD’s output was analyzed by a digital communication analyzer (DCA, Agilent 86100C with a 50 GHz 83484A module) to record the eye diagram. A voltage source (VS) adjusted the MZI’s heaters, enabling precise tuning of its operational state.

Figure \ref{fig.eye}(b) presents optical eye diagrams measured when the laser's wavelength was set to align  the MZI’s center wavelength (1541.1 nm) for both the "on" and "off" states. Plots (i) and (ii) correspond to PRBS data rates of 14 Gbps and 29 Gbps, respectively. The "on" and "off" voltages were approximately 1.77 V and 2.97 V, corresponding to the MZI’s maximum and minimum transmission at the center wavelength. As expected, the eye diagrams transition from fully open to completely closed when the MZI is switched from the "on" to the "off" state for both data rates.

To further validate the channel-selective blocking/passing functionality, signal wavelengths outside the MZI’s operational band (1539.5 nm and 1542.5 nm) were tested, as shown in Figs. \ref{fig.eye}(c) and \ref{fig.eye}(d). In these cases, the eye diagrams remain fully open regardless of the MZI’s state, confirming the device’s wavelength-selective behavior.

\section{CONCLUSION}
In conclusion, we have developed a narrowband MZI enabling  precise transmission control within a targeted wavelength band while maintaining out-of-band transparency.  This device is constructed using a novel dual-mode waveguide system (supporting TE$_0$ and TE$_1$ modes), incorporating 50\%-reflective AM-WBGs combined with asymmetric Y-branches to realize equivalent narrowband 1$\times$2 or 2$\times$2 couplers.
Experimental results have demonstrated a broad extinction ratio tuning range from 0 dB to $\sim$30 dB over a bandwidth of $\sim$2.5 nm, with successful simultaneous independent tuning of both wavelength and extinction ratio. Furthermore, characterization of cascaded systems has  validated their ability to precisely control intensity at individual wavelengths without inducing crosstalk. Finally, the narrowband MZI has been successfully demonstrated as a tunable, channel-selective optical blocker/passer for high-speed communication systems.

Compared to previous grating-assisted CDC designs, our AM-WBG-based approach fundamentally overcomes the bandwidth limitations imposed by unintended intra-waveguide coupling bands. Moreover, the single-waveguide-grating structure of AM-WBGs enhances reliability of both fabrication  and spectral control, while also enabling significant device miniaturization through spiral configurations, as  demonstrated previously \cite{Simard13, chen2015spiral, Zou16, Ma18, Cheng18}. 
These combined advantages establish the proposed narrowband MZI as a promising, scalable fundamental building block for advanced arbitrary spectral shaping and manipulation applications.

\section*{Funding}
National Natural Science Foundation of China (62105089); Science and Technology Innovation Yongjiang 2035 Key Research and Development Program(2025Z089).

\section*{Disclosures}
The authors declare no conflicts of interest.

\bibliography{sample}

\end{document}